\documentclass{article}
\usepackage{graphicx} 
\usepackage{amsmath}
\usepackage{amsfonts}
\usepackage{threeparttable} 
\usepackage{url}
\usepackage{float}
\title{Variational free complement method with Gaussian-expanded complement functions: convergence with fixed Gaussian expansion length}
\author{Cong Wang \thanks{PO Box 26 Okemos, MI, 48805 (USA); congwang.webmail@gmail.com}}
\date{}

\newcommand{\G}{\mathrm{G}}

\begin{document}

\maketitle

\begin{abstract}
For the free complement theory with Gaussian-expanded complement functions, the energy convergence of $n_\G = \mathrm{constant} < \infty, n\rightarrow\infty$ is discussed, where $n_\G$ is the number of the Gaussian functions in the STO-$n$G expansion.

\end{abstract}

\section{Introduction} \label{sec_intro}
A general method for obtaining numerically exact solutions of the time-independent Schr\"{o}dinger equation with feasible computational cost would provide reference values of many computational approaches \cite{shiozaki2009higher,Valeev_PRF_2009_46811G6,bubin2013born,mitroy2013theory,jiang2025neural} and data for machine learning \cite{slootman2024accurate,scherbela2024towards}. The free complement theory \cite{nakatsuji2004scaled,nakatsuji2005general,nakatsuji2012discovery} is a framework to these goals. (Recently, the theory has been renamed to free complete-element \cite{nakatsuji2024accurate, nakatsuji2024exact, nakatsuji2026solving}. Nevertheless, the completeness of basis set expansions can be proven \cite{klahn1977convergence_1,klahn1977convergence}. The proof of the free complete-element approach is heuristic \cite{nakatsuji2004scaled,nakatsuji2005general} and an counterexample has been reported  \cite{wang2025variational}, to the best of the present author's understanding and knowledge. Thus, we use the name free complement in the present work.) 

We have introduced the Gaussian-expanded complement approaches to circumvent the integral difficulties \cite{wang2025variational} and the hierarchical decontraction to mitigate the exponential complexities of the number of the complement functions \cite{wang2026free}.  For the ground state of the helium atom, both approaches can provide energies agree with the numerically exact energies within $1 \times 10^{-6}$ atomic units (a.u.) \cite{wang2025variational,wang2026free}. In present work, a.u. are used by default.

There are two parameters in the Gaussian-expanded complement function methods: order of the free complement method, $n$, and the length of the Gaussian expansion $n_\G$, besides the threshold of the overlap selection, descending/ascending in arrange the Gaussian functions, and if we regard two functions are the same within some numerical threshold (e.g., relative difference below $1 \times 10^{-9}$) \cite{wang2025variational,wang2026free}. For a Slater-type initial wavefunction and $1-e^{-\gamma r}$ type $g$ function \cite{nakatsuji2022accurate,nakatsuji2024accurate,nakatsuji2024exact}, if $n \rightarrow \infty$, $n_\G \rightarrow \infty$, and relax other numerical thresholds, we expect to obtain the ground state energies as from the methods without the Gaussian expansions. Furthermore, the completeness of these Gaussian functions provides flexibility to span a general function \cite{klahn1977convergence_1, wang2026free}.

For a fixed and finite $n_\G$ from a Slater-type initial wavefunction, the Gaussian expansion of the initial wavefunction will generate a non-zero lower bound of the exponents of the Gaussian functions in all orders of the free complement method  \cite{wang2026free}. Since the Gaussian wavefunctions have faster decays than the exact wavefunction \cite{morrell1975calculation,hoffmann1977schrodinger,hoffmann1979lower,hoffmann1980lower,katriel1980asymptotic,ahlrichs1981bounds,simon1982schrodinger,froese1983exponential,ahlrichs1989basic,fournais2008local,agmon2014lectures},  it leaves a question, for a fixed $n_\G$ and $n \rightarrow \infty $, if the energy converges to the eigenvalue of the Hamiltonian.

For the hydrogen ground state with $\psi_0 = e^{-\zeta r}$, $g=1 - e^{- \gamma r}$, and $\gamma = 1 - \zeta$, if $\zeta = 1$, the complement function will only have the initial wavefunction $\psi_0$ by the p-alone method \cite{nakatsuji2022accurate}. Hence, $n_\G = 1$ will not converge to the ground state energy. Notice in the present work, we omit the spin function. 

It would be more meaningful to consider the possibility that the number of the complement functions, $M_n$, could increase. Additionally, it has been noticed that the free complement method with the $1- e^{-\gamma r}$ type $g$ function can generate terms including $\frac{1- e^{-\gamma r}}{r}$. These terms present severe linear dependency with inefficient energy improvement \cite{nakatsuji2024accurate}. They are not included in the standard free complement calculations \cite{nakatsuji2024accurate}, nor the previous \cite{wang2025variational,wang2026free} and the present Gaussian-expanded-complement approaches. The possible role of kinetic operator will be discussed in the Summary and outlook section of the present work.

Therefore, the question (Q)  of the present work would be:
\begin{itemize}
    \item [Q:]  for a fixed $n_\G$, $n \rightarrow \infty$, $M_n \rightarrow \infty$, within the p-alone approach, and without the $\frac{1- e^{-\gamma r}}{r}$-type terms, will the methods converge to the exact eigenenergy? 
\end{itemize}

One may infer the answer (A) to Q is no, due to lack of diffuse function. Notice one sufficient condition of the Gaussian basis completeness requires an accumulative point $\alpha$ satisfies, $0 < \alpha  < \infty$ \cite[Theorem 5]{klahn1977convergence}. This is similar to a single Gaussian-type initial wavefunction with the $1-e^{-\gamma r}$ type $g$ function  \cite{wang2026free} when $n_\G \rightarrow \infty$.

To further address Q, the  present work shall use the non-relativistic electronic hydrogen ground state as the example.

The present work is organized as following. In Section \ref{hydrogen}, the results of the ground state of the hydrogen atom are presented and discussed for $n_\G=1, n\rightarrow \infty$. For $n_\G=1$, the decontracted and hierarchically contracted approaches are the same. In Section \ref{hydrogen_2}, a numerical study for $e^{-\alpha r^2}$ initial wavefunction with $1-e^{-\gamma r}$ $g$ function as $n_\G \rightarrow \infty$ of the hydrogen ground state is presented. This is to compare with Section \ref{hydrogen}. This also provides a more definitive answer for the energy convergence with a Gaussian initial wavefunction in the previous work \cite[Section 2.2]{wang2026free}. 
In Section \ref{method}, the main software and libraries used in the present work are described. In Section \ref{summary_outlook}, summary and outlook are provided. In Appendix, a comparison of energy convergences among even-tempered basis sets \cite{kutzelnigg2013expansion}, STO-$n$G \cite{hehre1969self,o1966gaussian,fernandez1988accurate}, decontracted STO-$n$G basis \cite{hesselmann2005efficient} sets are provided to support the exponential-type convergence for the STO-$n$G basis set \cite{hehre1969self,o1966gaussian,fernandez1988accurate} with and without contractions.

\section{Hydrogen atom ground state with $n_\G = 1$ and $n\rightarrow \infty$} \label{hydrogen}
\subsection{Theoretical analysis}

For the hydrogen ground state with the p-alone method \cite{nakatsuji2022accurate}, the Slater initial wavefunction, and the $1-e^{-\gamma r}$ type $g$ function \cite{nakatsuji2022accurate,nakatsuji2024accurate,nakatsuji2024exact}, the complement functions before the Gaussian expansion, $\{\phi_k\}$, at order $n$ can be  written as 
\begin{align}
  \phi_k &= g^k \psi_0,  k = 0, 1, \cdots, n  \\
    \psi_0 &= e^{-\zeta r} \\
  g &= 1 - e^{-\gamma r},  \gamma = 1 - \zeta
\end{align}

For $n_\G=1$, the decontracted \cite{wang2025variational}  and hierarchically contracted \cite{wang2026free}  approaches become the same:

\begin{align}
    e^{- (\zeta + k \gamma) r} &\rightarrow e^{-\alpha^{(k)} r^2}  \label{1g} \\
  \alpha^{(k)} &=  \alpha^{(0)}  \left( 1 + k\frac{\gamma}{\zeta} \right)^2  \label{alpha_series} \\
  \{ \phi_k | k = 0, 1, \cdots, n  \} &\rightarrow \{  e^{-\alpha^{(k)} r^2} | k = 0, 1, \cdots, n  \} \label{subsitution}
\end{align}
At order 0,  the exponent of the STO-1G expansion for $\zeta=1$ is: $\alpha^{(0)}   = \textrm{0.270 950}$ \cite{riera1972gaussian}. Eq. \eqref{alpha_series} is obtained by the scaling relation,  $ ( \zeta_1 / \zeta_2 )^2 =\alpha_1 / \alpha_2$, for STO-$n$G basis sets \cite{o1966gaussian}.

The complement functions \eqref{alpha_series} and \eqref{subsitution} are outside the known sufficient conditions of the Gaussian basis functions \cite[Theorem 5]{klahn1977convergence_1}.

The M\"untz–Sz\'asz theorem and its generalizations 
\cite{klahn1977convergence_1,klahn1985generalization,klahn1985convergence,hill1998completeness} may provide certain insights into the convergence of the complement functions \eqref{1g} and the associated energies. The theorem states that a basis $\{ x^{\lambda_k} \}_{k=1}^\infty$ with $\lambda_k \in \mathbb{R}, \lambda_k> -1/2$, and $\lambda_i \neq \lambda_j, \forall i \neq j$, is complete in $L^2([0,1])$, if and only if  \cite{klahn1985generalization}
\begin{align}
  \sum_{k=1}^{\infty} \frac{ 1 + 2 \lambda_k}{1 + \lambda_k^2} = \infty \label{ms_theorem} 
\end{align} 

The generalization of the sufficiency of the M\"untz–Sz\'asz theorem  for basis completeness and thus wavefunction convergence in $L^2(\mathbb R^3)$ space is similar to Eq. \eqref{ms_theorem} \cite{klahn1985convergence}. In $W^{(1)}_2(\mathbb R^3)$ space, the summation related to Eq. \eqref{ms_theorem} needs to be restricted to a finite exponent for the sufficiency \cite{klahn1985convergence}. Notice the basis completeness in $W^{(1)}_2(\mathbb R^3)$ space is sufficient for the energy convergence \cite[lemma 7, Theorem 7, and Eq. (21)]{klahn1977convergence_1}. The necessity of the energy convergence has been obtained for the non-degenerate energy eigenstate without spin function, which only requires the energy eigenfunction converged by the basis in the sense of the $A$-norm \cite[Theorem 6]{klahn1977convergence_1}. 

The necessary condition of the M\"untz–Sz\'asz theorem in $L^2(\mathbb R^3)$ space is unknown \cite{klahn1985convergence}.

Then, we can form $\{ (e^{-r^2})^{\alpha^{(k)}} \}$ that aligns with Ref. \cite{klahn1985generalization}, up to a possible shift of the index $k\rightarrow k + 1$. For a large $k$, the term in left-hand side of Eq. \eqref{ms_theorem} becomes
\begin{align}
     \frac{ 1 + 2 \lambda_k}{1 + \lambda_k^2} \rightarrow \frac{2}{\alpha^{(0)} (\gamma/\zeta)^2}  \frac{1}{  k^2 } \label{k_large}
\end{align}
Hence, the associated summation will be finite. Notice $M_n \rightarrow \infty$ in Q requires $\gamma \neq 0$ in \eqref{k_large}.

This suggests the sufficient conditions \cite{klahn1985convergence} of the completeness are not fulfilled. Though the necessary conditions in $L^2(\mathbb R^3)$ have not been obtained \cite{klahn1985convergence}, \eqref{k_large} indicates this is unlikely to be satisfied (the angular functions are constants for the ground state of the hydrogen atom). Since  convergence in $W_2^{(1)}$ space implies convergence in $L^2$ space   \cite[Figure 1]{klahn1977convergence}, by contraposition, the convergence in $W_2^{(1)}$ space via \eqref{k_large} is unlikely. 
 
However, the M\"untz–Sz\'asz  theorem and its generalizations  \cite{klahn1985generalization,klahn1985convergence} are regarding the convergence to any function in a given functional space. As suggested in the necessary condition of the energy convergence \cite[Theorem 6, also the following paragraph]{klahn1977convergence}, this does not exclude the possibility that a specific basis, as the free complement approach \cite{nakatsuji2004scaled,nakatsuji2005general,nakatsuji2012discovery}, to capture a specific state, the energy eigenstate of the Hamiltonian of the free complement method, and its energy. If this were true, the scenario would be far-reaching. (For a spinless formulation, the ground state of the hydrogen atom is non-degenerate. There can be trivial special examples via symmetry. Namely, for the $S$-state of the hydrogen atom, no need to include $p$-type basis functions. The $W_2^{(1)}(\mathbb R^3)$  sufficient completeness conditions require $\forall l \geq 0$ \cite{klahn1977convergence_1,klahn1985generalization}. We do not consider this possibility.)

Notice the free complement method has included functions lack the cusp singularities  \cite{kato1957eigenfunctions,pack1966cusp}  at finite order, as the "approximate" class \cite{nakatsuji2022accurate}. This can include the Gaussian-expanded complement functions \cite{wang2025variational,wang2026free}. 

\subsection{Numerical analysis}
In Table \ref{tab1},  a list of results from the Rayleigh-Ritz variational calculations of the complement functions via \eqref{1g} is reported.  Since these are baseline-type calculations, no selection based on the overlap matrices \cite{von2008trapped,von2009correlated,rakshit2012hyperspherical,mitroy2013theory,Kalaee2014,Mosegaard2018,moriya2023novel,coomar2022quantum,wang2025variational} nor removal of basis function by the canonical orthogonalization \cite{szabo1996modern} is performed. We have used 1200 decimal precision for $0 \leq n \leq 1000$ and 1500 decimal precision for $n=1200$, see Section \ref{method} for the technical descriptions. To justify the setting of numerical precision, the minimum eigenvalue of the overlap matrix, $s_{\mathrm{min}}$, is reported in  Table \ref{tab1}.

The energies in Table \ref{tab1} suggest a slow convergence towards an value above the exact solution, -0.5 a.u.. To extrapolate the limit of $n\rightarrow \infty$, we use the fitting with the form $E = A + B /M_n^{C}$.  $M_n\geq 100$ for the data in Table \ref{tab1} was adopted to suppress possible higher-order contributions, e.g., $B' /M_n^{C'}$ with $C'>C$  \cite{bromley2007convergence} and reduce the complexities of multiple solutions with several-term nonlinear fittings. The result of the fitting is $A = -\textrm{0.499 800 472} \pm 6 \times 10^{-9}$, $B = \textrm{0.001 246} \pm 7  \times 10^{-6}$, $C=  1.012 \pm 1\times 10^{-3}$, with $R^2 \approx  \textrm{0.999 999 3}$.

This rate of convergence does not invalid the exponent-type convergence \cite{klopper1986gaussian, kutzelnigg1994theory,kutzelnigg1996convergence,Kutzelnigg2011OWR,kutzelnigg2012rate,bakken2004expansion,mckemmish2012gaussian,kutzelnigg2013expansion,bachmayr2014error,shaw2020completeness,kutzelnigg1996convergence,wang2013rates} of Gaussian basis functions. The latter typically refers to uncontracted even-tempered \cite{kutzelnigg1994theory,kutzelnigg1996convergence,Kutzelnigg2011OWR,kutzelnigg2012rate,mckemmish2012gaussian,kutzelnigg2013expansion,bachmayr2014error,shaw2020completeness,wang2013rates}, uncontrated optimized \cite{klopper1986gaussian,bakken2004expansion}, and contracted basis sets \cite{jensen1999basis,jensen2005estimating,karton2006comment}. In Appendix, we provide a comparison between the even-tempered and STO-$n$G basis sets.

\begin{table}[H]
\caption{Results of the variational calculations via the Gaussian-expanded complement functions for the ground state of the hydrogen atom via Eqs. \eqref{alpha_series} and \eqref{subsitution}.  $\psi_0 = e^{-\zeta r }$, $\zeta = 0.5$ \cite{nakatsuji2009does}. $g=1-e^{-\gamma r}, \gamma=0.5$, $n_\G=1$. $n$, $M_n$,  $s_{\mathrm{min}}$, and $E$  are denoted to the order of the free complement method, the number of the complement functions, the minimum eigenvalue of the overlap matrix and the electronic energy \cite[p. 43]{szabo1996modern}, respectively. $s_{\mathrm{min}}$ and $E$ are rounded to even \cite{python_format,python_round,goldberg1991every}. }

\begin{threeparttable}
\begin{tabular}{llll}
\hline
$n$ &  $M_n$ &   $s_{\mathrm{min}}$ & $ E$   \\ \hline
      0   & 1     & 1.0$\times10^{0}$      & -0.313 715 465 448 \\
      10   & 11     & 3.2$\times10^{-9}$      & -0.499 346 141 271 \\
      20   & 21     & 4.0$\times10^{-19}$      & -0.499 718 224 594 \\
      50   & 51     & 9.2$\times10^{-50}$      & -0.499 776 603 096 \\
      100   & 101     & 1.6$\times10^{-101}$      & -0.499 788 810 231 \\
      200   & 201     & 1.3$\times10^{-205}$      & -0.499 794 667 468 \\
      300   & 301     & 6.3$\times10^{-310}$      & -0.499 796 610 448 \\
      400   & 401     & 2.3$\times10^{-414}$      & -0.499 797 581 416 \\
      500   & 501     & 7.5$\times10^{-519}$      & -0.499 798 163 947 \\
      1000   & 1001     & 1.2$\times10^{-1041}$      & -0.499 799 329 090 \\
      1200   & 1201     & 7.3$\times10^{-1251}$      & -0.499 799 523 311 \\
\hline
\end{tabular} \label{tab1}
\end{threeparttable}
\end{table}

\section{Hydrogen atom ground state with $\psi_0 = e^{-\alpha r^2}$, $n_\G \rightarrow \infty$, and $n\rightarrow \infty$} \label{hydrogen_2}
For comparison, we perform variational computations for the ground state of the hydrogen atom with a Gauss-type initial wavefunction, $\psi_0 = e^{-\alpha r^2}$. The lower bound of the Gaussian exponents will be determined by the initial wavefunction \cite{wang2026free}, regardless of the possible value of $n_\G$. To the contrary of Section \ref{hydrogen}, with this initial wavefunction and the limit $n_\G \rightarrow \infty$, we may expect an arbitrary number of exponents at the interval $(\alpha - \delta, \alpha + \delta)$.

To resemble this limit, we may use the $1-e^{-\gamma r}$ type $g$ function without the Gaussian expansions \cite{wang2025variational,wang2026free}.

After linear combinations as reorganization, we consider the complement functions $\{ \phi_k\}$ at order $n$:
\begin{align}
  \phi_k &= e^{- k \gamma r} \psi_0, k=0,1,\cdots, n;  \gamma = 1  \label{cf_form_2}   \\
   \psi_0 &= e^{-\alpha r^2}, \alpha =0.0625 \label{cf_form_2_psi0} \,\, \cite{nakatsuji2009does}
\end{align}

Using \cite{kikuchi1954gaussian,shavitt1962multicenter}
\begin{align}
e^{- k \gamma r} e^{-\alpha r^2} = \frac{k\gamma}{ 2\sqrt{\pi}}  \int_0^\infty s^{-3/2} e^{-\frac{ (k\gamma)^2}{4s} - (s+\alpha) r^2 } ds
\end{align}
we may expect $\alpha$ is similar to an accumulative point by considering $s\rightarrow 0, s^{-3/2} \rightarrow \infty,  k=1,2,\cdots$. 

Alternatively, the complement functions \eqref{cf_form_2} may be viewed as similar to $\{ r^k e^{-\alpha r^2} \}$ \cite{nakatsuji2009does}. The completeness of basis functions and  has been proved in Ref. \cite[Theorem 5]{klahn1977convergence} and demonstrated numerically in Ref. \cite{nakatsuji2009does}. 

The numerical results of the variational calculations for the complement functions Eqs. \eqref{cf_form_2} and \eqref{cf_form_2_psi0} are in Table \ref{tab2}.

\begin{table}[H]
\caption{Results of the variational calculations via the Gaussian-Slater mixed complement functions, Eqs. \eqref{cf_form_2} and \eqref{cf_form_2_psi0}, as the Gaussian-expanded complement functions with $n_\G \rightarrow \infty$, for the ground state of the hydrogen atom. $\psi_0 = e^{-\alpha r^2 }$, $\alpha = 0.0625$ \cite{nakatsuji2009does}. $n$, $M_n$,  $s_{\mathrm{min}}$, and $E$  are denoted to the order of the free complement method, the number of the complement functions, the minimum eigenvalue of the overlap matrix and the electronic energy, respectively. $s_{\mathrm{min}}$ and $E$  are rounded to even. }

\begin{threeparttable}
\begin{tabular}{llll}
\hline
$n$ &  $M_n$ &   $s_{\mathrm{min}}$ & $ E$   \\ \hline
      0   & 1     & 1.0$\times10^{0}$      & -0.305 192 280 401 \\
      10   & 11     & 5.9$\times10^{-12}$      & -0.499 977 220 797 \\
      20   & 21     & 3.0$\times10^{-26}$      & -0.499 998 758 185 \\
      50   & 51     & 8.2$\times10^{-71}$      & -0.499 999 894 771 \\
      100   & 101     & 2.5$\times10^{-146}$      & -0.499 999 935 676 \\
      200   & 201     & 2.2$\times10^{-298}$      & -0.499 999 980 254 \\
      300   & 301     & 7.0$\times10^{-451}$      & -0.499 999 991 734 \\
      400   & 401     & 1.5$\times10^{-603}$      & -0.499 999 995 783 \\
      500   & 501     & 2.5$\times10^{-756}$      & -0.499 999 997 562 \\
      1000   & 1001     & 8.0$\times10^{-1521}$      & -0.499 999 999 604 \\
      1200   & 1201     & 9.1$\times10^{-1827}$      & -0.499 999 999 760 \\
\hline
\end{tabular} \label{tab2}
\end{threeparttable}

\end{table}

Though the energy convergence in terms of $E$ and $M_n$ is slower than the $\{r^k e^{-\alpha r^2} | k=0,1,2,\cdots,n \}$ form of complement functions \cite[Table III]{nakatsuji2009does}, it appears to converge to the exact value, $-0.5$ a.u..

\section{Methods and implementations} \label{method}
For Section \ref{hydrogen} and Appendix, the integrals are evaluated by the formulae in Ref. \cite{mitroy2013theory} of the one-electron hydrogen system. For Section \ref{hydrogen_2}, the formula with the parabolic cylinder function is based on \cite[(19.5.3)]{abramowitz1964handbook} \cite[(12.5.1)]{NIST:DLMF} (also see Ref. \cite{ammar2023use}) is used.

The program language Python  \cite{van1991interactively} version 3.12.3 with third-party libraries cachetools \cite{cachetools} version 7.0.3 (also in the previous works \cite{wang2026free}), mpmath version 1.3.0 \cite{mpmath}, SymPy \cite{10.7717/peerj-cs.103} version 1.14.0 , SciPy \cite{2020SciPy-NMeth} version 1.16.0, and matplotlib \cite{Hunter:2007} version 3.8.0 have been used for reloading computed integrals, multi-precision computations, symbolic computations, data fitting (by the curve\_fit function \cite{fitted_curve}), and plotting respectively. The pcfu function in the mpmath library \cite{pcfu} has been used for the parabolic cylinder function.

In the present work, the reloading integrals are based on a preliminary implementation of the $N$-electron explicitly correlated Gaussian integrals in Ref. \cite{mitroy2013theory} than two-electron systems of the previous works \cite{wang2025variational,wang2026free}. In the current implementation, only the overlap integrals have been reloaded. 

The Julia program language \cite{Julia-2017} version 1.11.6 has been used in solving the generalized eigenvalue problem \cite{wang2025variational} from the Rayleigh-Ritz variations with the BigFloat type for arbitrary numerical precision computations. The Julia third-party library GenericLinearAlgebra \cite{GenericLinearAlgebra} version 0.3.18 has been used in matrix multiplications and eigenvalue solutions. The Julia third-party library JSON \cite{json} version 0.21.4 has been used in data transfer between Python and Julia (also in the previous works \cite{wang2025variational, wang2026free}).

The decimal precisions of mpmath and BigFloat are set to be the same. They are 1200 and 1500 for $0 \leq n \leq 1000$ and $n=1200$, respectively from the data in Table \ref{tab1}.
For Table \ref{tab2}, 1500 and 2000 decimal precisions are used for $0 \leq n \leq 500$ and $n=1000$, respectively. For the fitted function in Section \ref{hydrogen} and numerical data in Appendix including the computed values of $\alpha, \beta,\omega$ to compare Ref. \cite{kutzelnigg2013expansion}, Table \ref{tab_a1}, and Figure \ref{fig_a1}, 32 decimal precision is used. Nevertheless, SciPy and matplotlib are in double precision.   

In solving the generalized eigenvalue problem by the canonical orthogonalization method \cite{szabo1996modern,wang2025variational}, we set the threshold of removing basis functions associated to smaller eigenvalue of the overlap matrix as $1\times 10^{-d}$ for $d$-digit numerical precision. No complement function was removed. 

The uncertainties of the fitted data in Section \ref{hydrogen_2} is one standard deviation \cite{fitted_curve}. The initial guess for the parameters $A, B$, and $C$ are the energy with the largest $M_n$ (1201 in Table \ref{tab1}), 0.01, and 1.0, respectively. 

The cursor system \cite{cursor} has been used in generating the Python code. 

\section{Summary  and outlook}  \label{summary_outlook}
By numerical evidence and theoretical analysis, the tentative A for the Q in Section \ref{sec_intro} is
\begin{itemize}
    \item [A:] no.
\end{itemize}
Additional remarks are:
\begin{itemize}
    \item [(i)] The Gaussian-expanded free complement methods \cite{wang2025variational,wang2026free} with the p-alone approach \cite{nakatsuji2022accurate} are unlikely to have a stronger computational capacity to capture the ground state energy than the basis completeness for a general function via the M\"untz–Sz\'asz theorem and its generalizations \cite{klahn1985convergence,klahn1985generalization};
    \item [(ii)] Lack of diffuse function does not necessarily imply energy non-convergence, it can be compensated similarly by an accumulative point \cite[Theorem 5]{klahn1977convergence};
    \item [(iii)] This work does not necessarily imply a disadvantage of the Gaussian-expanded free complement methods. In practical calculation, finite $n_\G>1$ and $n$ will be chosen for the target accuracy;       
    \item [(iv)] The appeared non-convergence of energy in Table \ref{tab1} may be cured by optimizing the exponents other than order 0. This would make the Gaussian-expanded free complement methods \cite{wang2025variational,wang2026free} similar to the explicit correlated Gaussian methods \cite{mitroy2013theory,bubin2013born}. This is computationally non-preferable;   
    \item [(v)] One may try Gaussian-type $g$ function, e.g., $1 - e^{-\gamma r_{}^2} $ form $g$ function \cite[Eq. (19d)]{nakatsuji2022accurate} instead of Slater-type initial wavefunction and $g$ functions with Gaussian-expanded complement functions. The numerical examples with optimized exponents are below 1 kcal mol$^{-1}$ or 0.1 kcal mol$^{-1}$ \cite[$G_{12}$ or $G_{27}$ in Table II]{nakatsuji2022accurate}.  
    Without optimizing the exponents other than order 0, the denominator at the right-hand side of Eq. \eqref{k_large} would be at linear order of $n$. This will fulfill the sufficient condition of the generalized M\"untz–Sz\'asz theorem in $L^2(\mathbb R^3)$, but not in $W_2^{(1)}(\mathbb R^3)$ \cite{klahn1985generalization}. The latter requires the exponents smaller than an $l$-dependent constant \cite{klahn1985generalization}. In this instance, the convergence of energy without optimizing the exponents other than order 0 is an open question. The linear dependence is expected to be more severe than the Gaussian-expanded complement functions. 
    It is unlikely to be more effective than appropriate chosen $n_\G$ and $n$ in the Gaussian-expanded complement approaches;
    
    \item [(vi)] If one introduces the kinetic operator, as the p+k approach \cite{nakatsuji2004scaled,nakatsuji2005general,nakatsuji2022accurate}, on the Gaussian-expanded complement functions, this could generate $\{ r_{}^{k_{}} e^{-\alpha r_{}^2}|k_{}=0,1,2,\cdots,n\}$ complement functions \cite{nakatsuji2009does,kurokawa2023gaussian}. Notice a higher power of $k$ has the maximum absolute value of the amplitude at a longer range, thus this could possibly compensate the incompleteness in the aspect (i) of the present Section. One could further form Gaussian-expanded complement functions from expanding $\{ r_{}^{k_{}} e^{-\alpha r_{}^2}|k_{}=0,1,2,\cdots,n\}$. The Gaussian expansion for ${k_{}}=1$ \cite{kurokawa2023gaussian} may only be needed, since higher odd power expansion can be rearranged as higher angular momenta \cite{nakashima2020free,nakatsuji2020solving}. For example, in the decontracted approach \cite{wang2025variational}, $ (r_{}^2)^{\frac{k_{}-1}{2}}  r_{}^{1} e^{-\alpha r_{}^2}  \rightarrow \{ (r_{}^2)^{\frac{k_{}-1}{2}} e^{-\alpha^{(m) r^2}}| m=0,1,2,\cdots \} $, where $k=1,3,5,\cdots $.  This may reduce the need of a large $n_\G$ in many-electron computations in the hierarchical decontraction approach \cite{wang2026free}.
\end{itemize}

\appendix
\renewcommand{\thetable}{A\arabic{table}}
\renewcommand{\thefigure}{A\arabic{figure}}
\setcounter{table}{0}

\section*{Appendix: Comparison for the even-tempered and STO-$n$G 
variational energies for the hydrogen ground state} \label{sec_append}

In this Appendix, we present a comparison among even-tempered with local approximation \cite{kutzelnigg2013expansion}, even-tempered with global approximation \cite{kutzelnigg2013expansion},  STO-$n$G, and decontracted STO-$n$G basis sets for the hydrogen ground state energies in Table \ref{tab_a1}. This is to support that from the STO-$n$G expansion, exponential-type convergence can be expected. The results with the STO-$n$G basis set resemble the terms without $g$ function in the hierarchical decontraction scheme \cite{wang2026free}. The results with the decontracted STO-$n$G basis set resemble the decontracted method \cite{wang2025variational}. 

In addition, $\zeta = 1.24$ has also been use for hydrogen in the STO-$n$G basis set \cite{szabo1996modern}. In the present work, we use $\zeta = 1$.  In this Appendix, when mentioning the formulae from Ref. \cite{kutzelnigg2013expansion}, the notation $n$ in Ref. \cite{kutzelnigg2013expansion} has been replaced to $M_n$ for consistency with the present work.

\begin{table}[H]
\caption{Comparison for the even-tempered and STO-$n$G variational energies for the ground state of the hydrogen atom. Local and global correspond to the local and global approximation schemes in Ref. \cite{kutzelnigg2013expansion}, respectively.  $n_\G$ corresponds to the STO-$n$G basis set ($n_\G= 3,6$ \cite{hehre1969self}, $10$ \cite{o1966gaussian}, and $14$ \cite{fernandez1988accurate}).  Contracted and decontracted correspond to the original expressions of the STO-$n$G basis sets and the decontraction of them, respectively. $M_n$,  $s_{\mathrm{min}}$, and $E$  are denoted to the number of linear independent functions in the variational calculations, the minimum eigenvalue of the overlap matrix and the electronic energy, respectively.   $s_{\mathrm{min}}$  and $E$ are rounded to even. }

\begin{threeparttable}
\begin{tabular}{cccccccccc}
\hline
\multicolumn{4}{c}{local} & \multicolumn{4}{c}{global} \\
$M_n$ &  &   $s_\mathrm{min}$ &  $E$ & $M_n$ & &  $s_{\mathrm{min}}$ & $E$  \\ \hline
3 &  & 5.5$\times10^{-1}$ & -0.451 282 421  &3 &  & 3.6$\times10^{-1}$ & -0.480 586 023  \\
6 &  & 1.9$\times10^{-1}$ & -0.493 049 625  &6 &  & 7.7$\times10^{-2}$ & -0.499 294 960  \\
10 &  & 5.9$\times10^{-2}$ & -0.498 889 681  &10 &  & 1.4$\times10^{-2}$ & -0.499 957 798  \\
14 &  & 2.1$\times10^{-2}$ & -0.499 852 426  &14 &  & 3.6$\times10^{-3}$ & -0.499 997 695  \\
\hline
\multicolumn{4}{c}{contracted} & \multicolumn{4}{c}{decontracted} \\
$M_n$ & $n_\G$ & $s_\mathrm{min}$ &  $E$ &  $M_n$  &  $n_\G$ & $s_{\mathrm{min}}$ & $E$ \\ \hline
1 & 3 &  1.0$\times10^{0}$ & -0.494 91\tnote{a} & 3 & 3 & 1.6$\times10^{-1}$ &  -0.495 010 402 \\
1 & 6 &  1.0$\times10^{0}$ & -0.499 83\tnote{a} & 6 & 6 & 1.5$\times10^{-2}$ &  -0.499 827 137 \\
1 & 10 &  1.0$\times10^{0}$ & -0.499 998 358 & 10 & 10 & 4.0$\times10^{-3}$ &  -0.499 998 506 \\
1 & 14 &  1.0$\times10^{0}$ & -0.499 999 798\tnote{b} & 14 & 14 & 9.2$\times10^{-5}$ &  -0.499 999 798\tnote{c} \\
\hline
\end{tabular} \label{tab_a1}
    \begin{tablenotes}
      \item[a] Ref. \cite{hehre1969self}
      \item[b] -0.499 999 797 545
      \item[c] -0.499 999 797 548
    \end{tablenotes}
\end{threeparttable}

\end{table}

\begin{figure}[H]
  \centering
  \includegraphics[width=10.5cm]{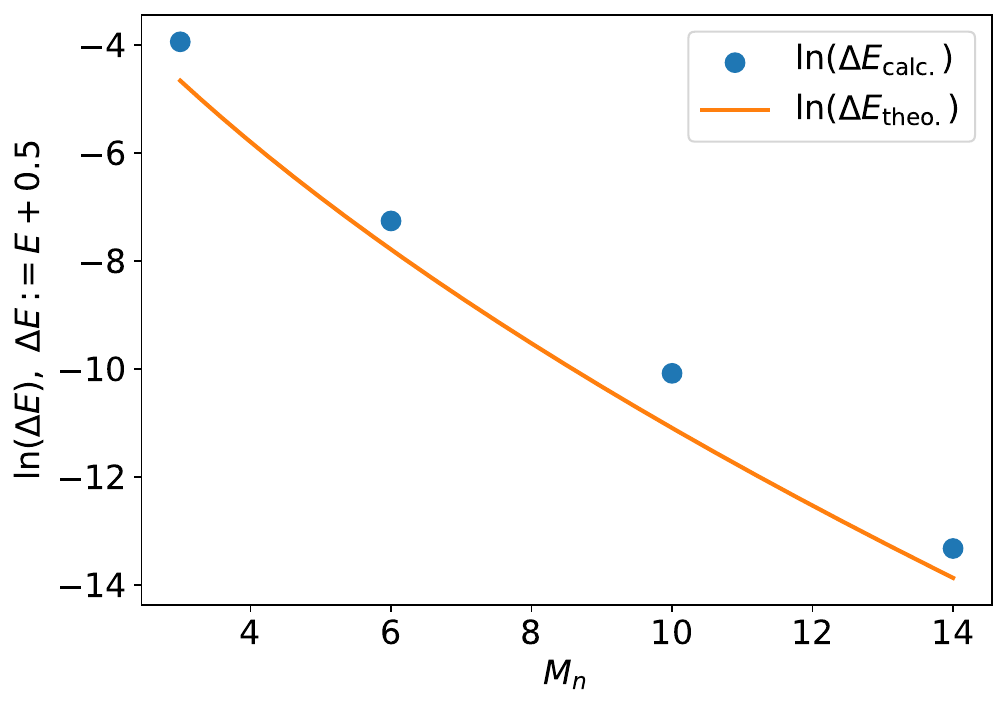}
  \caption{The plot $\ln \Delta E$  versus  $M_n$ for the variational energies from the global approximation of the even-tempered basis set \cite{kutzelnigg2013expansion} of the ground state of the hydrogen atom. $\Delta E_{\mathrm{calc.}}$ corresponds to the energies of $E$ under global in Table \ref{tab_a1}.  $\Delta E_{\mathrm{theo.}}$ corresponds to the theoretical error formula \cite[Eq. (95)]{kutzelnigg2013expansion} without $O(M_n^{-1/2})$, i.e., $\Delta E_{\mathrm{theo.}} :=  4 \cdot 3^{5/8} \sqrt{ 2 (8-5\sqrt{2} ) } \pi Z^2 M_n^{9/8} e^{-\pi \sqrt{3M_n}}$ with $Z=1$.  }
  \label{fig_a1}
\end{figure}

For the local even-tempered basis set in Table \ref{tab_a1}, we have implemented all terms up to the next leading order of $\alpha$ and $\beta$ in from Ref. \cite[Eqs. (43)-(48)]{kutzelnigg2013expansion}. For example, 
\begin{align}
    \alpha = \frac{Z^2}{2\pi \sqrt{2M_n}} \left[1 + \frac{\ln \left( 8 \pi^2 M_n^3 \right)}{4\pi \sqrt{2M_n}} \right] e^{\frac{1}{2} \left[ \pi \sqrt{ \frac{2}{M_n}} + \frac{\ln(2M_n)}{4M_n}\right]  } \label{alpha_nlo}
\end{align}
Hence, the expression $ \frac{Z^2}{2\pi \sqrt{2M_n}} \exp\left( \pi \sqrt{\frac{2}{M_n}}\right)$ for the leading order of $\alpha$ in Ref. \cite[Eq. (48)]{kutzelnigg2013expansion} should be  $ \frac{Z^2}{2\pi \sqrt{2M_n}} \exp\left( \frac{\pi}{2} \sqrt{\frac{2}{M_n}}\right)$. Small numerical discrepancies exist at $n=10$. Using the values in Ref. \cite[Table 1]{kutzelnigg2013expansion}, 0.098 667 and 4.392 31 for $\alpha$ and $\beta$ respectively, the formula $\omega=\alpha \beta^{n-1}$ \cite{kutzelnigg2013expansion} leads to 60 035.589 7 than 60 035.9.7 in Ref. \cite{kutzelnigg2013expansion}. With 32 decimal precision, the formulae including the next leading terms, Eq. \eqref{alpha_nlo} and Ref. \cite[Eqs. (43)-(48)]{kutzelnigg2013expansion}, and $Z=1$,  we obtain $\alpha \approx \textrm{0.098 666 629 2}$, $\beta \approx 	
\textrm{4.392 314 253 7}$. These lead to $\omega \approx \textrm{60 035.89}$ for $n=10$.  In addition, a square-root exponential energy-error formula has been obtained by numerical fitting  \cite[Eq. (61)]{kutzelnigg2013expansion}.

For the global approximated scheme \cite[Eqs. (91)-(94)]{kutzelnigg2013expansion}, no specific value of the numerical constant $d$ in $h_1$ nor the expression of $O(M_n^{-1/2})$ in $s_l$ for the next leading order were found in Ref. \cite{kutzelnigg2013expansion}. We are unable to reproduce the numerical values of $\alpha$ and $\beta$ in Table 2 of Ref. \cite{kutzelnigg2013expansion} without significant deviations. We only included the leading order \cite[Eqs. (96) and (97)]{kutzelnigg2013expansion} for $\alpha$ and $\beta$ for the global data in Table \ref{tab_a1}. The values have rather significant deviations  with the data in Table 2 in Ref. \cite{kutzelnigg2013expansion}. For example, for $M_n=10$, using the leading term expressions \cite[Eqs. (96) and (97)]{kutzelnigg2013expansion}, we obtain $\alpha \approx \textrm{0.025 783}, \beta \approx \textrm{3.149 20}$. The values with $M_n=10$ in Table 2 in Ref. \cite{kutzelnigg2013expansion} are 0.032 337 and 2.981 29 for $\alpha$ and $\beta$, respectively. Nevertheless, as provided in Figure \ref{fig_a1}, the computed energies align with the theoretical exponential-root error rate \cite[Eq. (95)]{kutzelnigg2013expansion}. 

For the same $M_n$, both contracted and decontracted energies are lower than local and global energies. These results provides numerical support that the complement functions constructed from STO-$n$G basis set can exhibit exponential-type variational energy convergence.

\bibliographystyle{pccp}
\bibliography{topic}

\end{document}